\documentclass[aps,pra,twocolumn,superscriptaddress,showpacs]{revtex4}
\usepackage{tipa}
\usepackage{pifont}
\usepackage{txfonts}
\usepackage{amssymb}
\usepackage{graphicx}

\begin{document}

\title{The scaling behavior of logarithmic fidelity in quantum phase transition in LMG model}
\author{Ching-Yee Leung}
\email{flyee.leung@polyu.edu.hk}
\affiliation{Department of Physics and ITP, The Chinese University of Hong Kong, Shatin,
Hong Kong, China}
\author{Ho-Man Kwok}
\affiliation{Department of Physics and ITP, The Chinese University of Hong Kong, Shatin,
Hong Kong, China}
\author{Shi-Jian Gu}
\affiliation{Department of Physics and ITP, The Chinese University of Hong Kong, Shatin,
Hong Kong, China}
\author{Hai-Qing Lin}
\affiliation{Department of Physics and ITP, The Chinese University of Hong Kong, Shatin,
Hong Kong, China}
\date{\today}

\begin{abstract}
In this paper, we explore the differences between classical logarithmic fidelity and quantum fidelity. The classical logarithmic fidelity is found to be always extensive while the quantum one manifests distinct size dependence in different phases. Illustrated by the anisotropic Lipkin-Meshkov-Glick model, we found numerically and analytically that the logarithmic fidelity scales like $N$ in the symmetry-broken phase and scales like $N^0$ in the polarized phase. The singular behavior around the critical point is also investigated.
\end{abstract}


\maketitle

%05.70.Fh    phase transitions: general studies
%03.67.Mn    entanglement production, characterization, and manipulation
%03.75.Gg    entanglement and decoherence Bose-Einstein condensates
%03.75.Hh    static properties of condensates, thermodynamical statistical
%            and structural properties
%75.10.jm    quantized spin models
%71.10.Fd    Lattice fermion models (Hubbard model, etc.)
%75.30.kz    magnetic phase boundaries (including magnetic transitions,
%            metamagnetism, etc)
%03.67.-a    quantum information
%64.60.-i    General studies of phase transitions (see also 63.70.+h
%Statistical mechanics of lattice vibrations and displacive phase transitions;
%for critical phenomena in solid surfaces and interfaces, and in
%magnetism, see 68.35.Rh, and 75.40.-s, respectively)
%75.10.-b General theory and models of magnetic ordering (see also
%05.50.+q Lattice theory and statistics)

%05.70.Fh, 03.67.Mn, 03.75.Gg, 03.75.Hh
% PACS Number

\section{Introduction}

Fidelity\cite{nilesen} is an information concept used to measure the
similarity of an input and its corresponding output (classical or quantum)
states of a channel. In quantum information theory, the fidelity concept is
hardly related to the size dependence because information science usually
focuses on quantum states (or information) carried by few-body systems.
Therefore, none of previous works on the fidelity in information science, as
far as we know, paid attention to the scaling behavior of the fidelity until
recent application of fidelity in quantum phase transitions(QPTs)
\cite{sachdev}.

Up to now, A lot of works \cite{ReviewFidelity} connecting QPTs
and the fidelity have been done. Traditionally, a QPT occurs when there is a significant
change of the system's physical quantity. But of the quantum ground state
concern, the fidelity measures the similarity between two quantum states
differed by a certain fixed value of a driving parameter. When these two
observations come together, the QPT can be observed as long as the fidelity
drops to zero. It can be understood as the state of the system undergoes
structural change from one phase to another, such that states from two phases
are orthogonal to each other \cite{HTQuan06,PZanardi06,HQZhou07}. Such an
interesting idea means the QPTs can be characterized by solely the quantum
state itself, without \textit{a priori} knowledge to the symmetry and order of
the system.

Since then, various fidelity-related measures have been put forward,
including the fidelity per site \cite{HQZhouFPS}, the fidelity
susceptibility \cite{PZanardi07FS,WLYou07FS}, thermal-state fidelity \cite%
{WLYou07FS,PZanardi07FS,PZanardi2007T,PZanardi062318,NPaunkovic08,HTQuan2009}%
, operator fidelity \cite{XWang012105}, and density-functional fidelity\cite%
{SJGuDFF}, which can be applied under different circumstances. Remarkably,
researches have been focusing on the scaling behavior of the above fidelity
measures, as the second-order QPTs occur only in the thermodynamics limit
\cite{LCVenuti07FS,SJGu07FS}. While the idea of critical
exponents in physical quantities has been well-established \cite%
{MAContinentino}, the above findings have also successfully related the
critical exponent of the fidelity to those of the physical quantities,
making the fidelity more physically meaningful.

However, it seems to us that the logarithmic fidelity, is presumably regarded an extensive quantity in relevant studies. Such
an idea is true for thermal state. As we will show below that the logarithmic
fidelity of two thermal states is proportional to the Helmholtz free energy,
which is an extensive quantity for a thermodynamic system. While dramatically the logarithmic fidelity of two quantum states have distinct sizse dependence in different phases, which is believed to be caused by various quantum correlations. To explore the difference between the thermal-state logarithmic fidelity and the ground-state logarithmic fidelity is the key motivation of the present work.

In this paper, we show that the scaling dependence of the logarithmic
fidelity may not always be universal for a thermodynamic system. The thermal-state logarithmic fidelity is extensive due to the extensibility of
the Helmholtz free energy. While the quantum ground-state logarithmic fidelity might
be either intensive, extensive, or superextensive, depending on the quantum
adiabatic dimension of the ground state. We take the anisotropic
Lipkin-Meshkov-Glick model (LMG model) \cite{LMG} as an example and show
that the logarithmic fidelity scales like $N$ in the model's symmetry-broken
phase and $N^0$ in the polarized phase.

This paper is organized as follows: In Sec. II, we introduce the definition
of fidelity and logarithmic fidelity for both thermal state and quantum state, then explain why
fidelity has different scaling behavior for both cases. In Sec. III, we take
the LMG model as an example, and explicitly show that the logarithmic
fidelity scales like $N$ in the symmetry broken phase and $N^0$ in the
polarized phase. In Sec. IV, we study numerically the critical properties of
the fidelity per site around the transition point. Finally, our conclusions
are given in Sec. V.

\section{Thermal-state fidelity and quantum-state fidelity}

Let a quantum-many system be characterized by a pure state $\left|\Psi(h)\rangle \right.$ labeled
the continuous variable $h$. In describing a QPT, the continuous variable is the driving parameter that
induces the QPT, very often it is the interaction strength, or external field strength. The project between two states of different
values of the variable $h$ and $h^{\prime }$ can be measured by the
fidelity, which is defined as
\begin{eqnarray}
F(h, h^{\prime }) = \left| \langle\Psi(h)|\Psi(h)\rangle \right|.
\end{eqnarray}
The fidelity is zero if two states are orthogonal, one if identical.

In Ref. \cite%
{HQZhou07}, the authors proposed the logarithm of the fidelity, namely the
fidelity per site to describe QPTs. It is given by
\begin{eqnarray}
d(h ,h ^{\prime }) &=& \mathop {\lim }\limits_{N \to \infty } F(h ,h
^{\prime})^\frac{1}{N}, \\
\ln d(h ,h ^{\prime }) &=& \mathop {\lim }\limits_{N \to \infty } \frac{1}{N}%
\ln F(h ,h ^{\prime }),
\end{eqnarray}
in which $N$ is the system size. Usually, fidelity scale like $d^N$ for
large scale $N$. Fidelity per site give a measurement of fidelity that is
independent of the system size.

The extension of fidelity to a mixed-state or matrix-product state makes use
of the density matrix $\rho(h)$ of the system, it is defined as
\begin{eqnarray}
F(h, h^{\prime }) = \mathrm{tr \sqrt {\rho^{1/2} (h)\rho (h^{\prime
})\rho^{1/2} (h)}.}
\end{eqnarray}
If two mixed states are diagonal in the same set of basis, the fidelity is
the trace of the product of the density matrices $F(h, h^{\prime }) =
\mathrm{tr \sqrt {\rho (h)\rho (h^{\prime })}}$. Such a definition allows
description of classical phase transitions, as long as the thermal state of
the system is obtained.

At finite temperatures, a thermal state is described by the density
matrix
\begin{eqnarray}
\rho (\beta ) = \frac{1}{Z}\sum\limits_n {e^{ - \beta E_n } \left| {\Psi _n }
\right\rangle \left\langle {\Psi _n } \right|}
\end{eqnarray}
where $\beta = T^{-1}$ denotes the inverse temperature, and the Boltzmann
constant $k$ is set to be one, $Z$ is the partition function of the system,
i.e. $Z \left( \beta \right) = \sum_n {e^{-\beta E_n}}$, and $\left| \Psi_n \rangle \right.$ is the energy 
eigenstate of the system's Hamiltonian. The thermal-state fidelity in the
parameter space of temperature (and thus $\beta$) is of the form
\begin{eqnarray}
F = \frac{Z\left(\frac{\beta_1+\beta_2}{2}\right)}{%
\sqrt{Z\left(\beta_1\right)Z\left(\beta_2\right)}},
\end{eqnarray}
with different temperature $T_i=\beta_i^{-1}$.

Let us take the logarithm of the thermal-state fidelity,
\begin{eqnarray}
\ln F  
=\ln Z\left( \frac{\beta _{1}+\beta _{2}}{2}\right) -\frac{1}{2}\ln
Z\left( \beta _{1}\right) -\frac{1}{2}\ln Z\left( \beta _{2}\right) .
\end{eqnarray}%
Notice that%
\[
G=-\frac{1}{\beta }\ln Z
\]
is the Helmholtz free energy, which is an extensive quantity, thus
logarithmic fidelity for thermal state should also be an extensive quantity. An other useful
quantum quantity, the fidelity susceptibility $\chi _{F}$ was also be derived to be
equal to $C_{v}/4\beta ^{2}$ \cite{PZanardi07FS,WLYou07FS}, that is
proportional to $N$.

On the other hand, the quautum-state logarithmic fidelity leads by the
fidelity susceptibility\cite{SJGu08},
\[
\ln F(h_{1},h_{2})=-\frac{(h_{1}-h_{2})^{2}\chi _{F}}{2}+\cdots ,
\]%
if $h_{1}$ and $h_{2}$ are close to each other. The ground-state fidelity
susceptibility has its own quantum adiabatic dimension in different phases, while the thermal-state logarithmic is lead by extensive term. Therefore, we speculate that the 
quautum-state logarithmic fidelity might not be always extensive. In the following
section, we are going to use the anisotropic LMG model as an example to
demonstrate the size dependence of logarithmic fidelity in both phases.

\section{The logarithmic fidelity in the anisotropic LMG model}

The Hamiltonian of the LMG model reads
\begin{eqnarray}
H_{\mathrm{LMG}} &=&-\frac{1}{N}\sum_{i<j}\left( \sigma _{x}^{i}\sigma
_{x}^{j}+\gamma \sigma _{y}^{i}\sigma _{y}^{j}\right) -h\sum_{j}\sigma
_{z}^{j}  \nonumber \\
&=&-\frac{1}{N}\left( {1+\gamma }\right) \left( {\mathbf{S}^{2}-S_{z}^{2}-N/2%
}\right) -2hS_{z}  \nonumber \\
&&-\frac{1}{{2N}}\left( {1-\gamma }\right) \left( {S_{+}^{2}+S_{-}^{2}}%
\right) ,  \label{eq:HLMG}
\end{eqnarray}%
where $\sigma _{\kappa }\,(\kappa =x,y,z)$ are the usual Pauli matrices, 
$S_{\kappa }=\sum_{j}\sigma _{\kappa }^{j}/2$ the collective operator, $%
\gamma \leq 1$ denotes the anisotropy parameter, and $h$ is the external magnetic field. In its isotropic case $(\gamma =1)$, it undergoes a first order QPT (ground state level crossing) at $|h|=1$. In our discussion, anisotropic case is concerned. In anisotropic case $(\gamma \neq 1)$, a second order QPT at $h=1$. The ground state of the system falls on $%
\mathbf{S}^{2}=\frac{N}{2}\left( \frac{N}{2}+1\right) $. The LMG model can
be used to describe the Bose-Einstein condensate and Josephson junctions.
The spectrum of the LMG model is recently reviewed by some sophisticated
analytic method, including the continuous unitary transformation \cite%
{SDusuel05} and the spin coherent state formalism \cite%
{PRibeiro07,PRibeiro08}, these enriched the understanding of the model, and
extended the study to a wide range of parameter values.

To study the behavior of logarithmic fidelity of the LMG model, we are going to discuss the polarized phase and symmetry-broken phase seperately.

\subsection{Polarized phase $(h >1)$}

In polarized phase, the usual treatment to obtain the solution of groundstate is by firstly, mapping the collective spin operator $\mathbf{S}$ into bosonic operators $a$ and $a^{\dagger}$ by Holstein-Primakoff transformation. Then $a$ and $a^{\dagger}$ are mapped into another pair of bosonic operator $b$ and $b^{\dagger}$ by standard Bogoliubov transformation. i.e,

\begin{eqnarray}
b &=& \cosh \frac{\theta}{2} a - \sinh \frac{\theta}{2} a^{\dagger} \nonumber \\
b^{\dagger} &=& \cosh \frac{\theta}{2} a^{\dagger} - \sinh \frac{\theta}{2} a. 
\end{eqnarray}

After adjusting the parameter $\theta$, the Hamiltonian is diagonalized to the following form while $\theta = \tanh ^{-1} \left( \frac{1-\gamma}{2h-1-\gamma}\right)$.

\begin{eqnarray}
H=h(N+1) +2\sqrt{(h-1)(h-\gamma)}\left(b^{\dagger}b + \frac{1}{2} \right)
\end{eqnarray}

The eigenstates are $ | n \rangle _b$, therefore the groundstate is $ | 0 \rangle _b$. We are interested in the fidelity $F(h,h') = | _b \langle 0(h) |  0(h')\rangle _b |$

\begin{figure}[h]
\includegraphics[width=8.5cm]{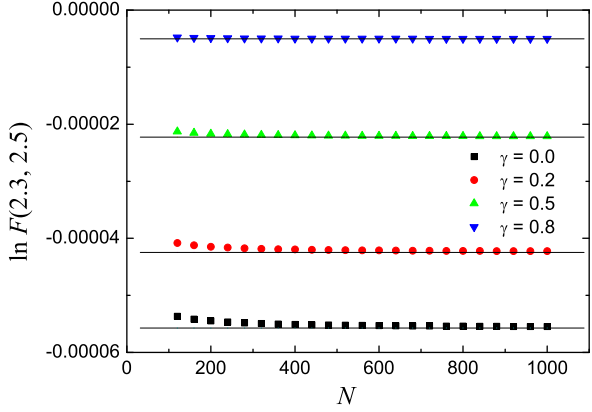}
\caption{(Color online)The dependence of logarithm of fidelity to the system size of the
LMG model in the symmetric phase. Solid lines represent the analytical
solution and dots represent the numerical results.}
\label{fig:hgone}
\end{figure}

In Fig. \ref{fig:hgone} we show the behavior of logarithm of fidelity in
the symmetry-broken phase ($h<1$). It shows clearly that logarithm of fidelity and
the system is in linear relation. i.e,
\[
\ln F\propto N,
\]%
such that
\[
d=\lim_{N\rightarrow \infty }F^{\frac{1}{N}}=\mathrm{constant.}
\]%
This observation means that the logarithmic fidelity is proportional to system size. However,
in the polarized phase ($h>1$), we obtain that the logarithmic fidelity is of different
dependence of system size. Refer to Fig. \ref{fig:hgone}, it tells us that
when $N$ is increased, the quantum logarithmic fidelity tends to a fix number. i.e,
\[
\ln F\propto N^{0}.
\]%
As its independence of system size, it is meaningless to measure $\frac{1}{N}%
\ln F$ is this phase. Its logarithmic fidelity always tends to zero while
system size increases.

Alternatively, we then confirm difference $N$ dependence by applying analytical methods to obtain explicitly the fidelity and its logarithm of the LMG model. By mapping the Hamiltonian into the bosonic operator and then diagonalizing it by standard Bogoliubov transformation, the Hamiltonian in the polarized phase ($h>1$) becomes

\[
H=-h(N+1)+2\sqrt{(h-1)(h-\gamma )}\left( b^{\dagger }b+\frac{1}{2}\right) ,
\]%
where the eigenstates are $\left\vert n\right\rangle _{b}$, and $b^{\dagger
}b\left\vert n\right\rangle _{b}=n\left\vert n\right\rangle _{b}$.
Therefore, we can start with the ground state
\[
b\left\vert \psi _{o}\right\rangle _{b}=0
\]%
where the relation between $a$ and $b$ is given by Bogoliubov transformation
that%
\[
b=\mathrm{cosh\frac{\theta }{2}a-sinh\frac{\theta }{2}a^{\dagger }}
\]
where $\theta $ is a function of $h$. It is possible to express the ground state
as the combination of $\left\vert m\right\rangle _{a}$, which is the
eigenstates of $a^{\dagger }a$,where $a^{\dagger }a\left\vert m\right\rangle
_{a}=m\left\vert m\right\rangle _{a}$, i.e,
\[
\left\vert \psi _{o}\right\rangle _{b}=\sum_{m=0}^{\infty }c_{m}\left\vert
m\right\rangle _{a},
\]%
where $c_{m}$ is the coefficient and it is then solved and normalized by $%
\langle \Psi _{o}|\Psi _{o}\rangle _{b}=1$ . The result shows,

\[
\left\{ {\begin{array}{*{20}c} {c_{2k} = \left(1 - \tanh ^2 \frac{\theta
}{2}\right)^{1/4} \tanh ^k \frac{\theta }{2}\sqrt {\frac{{(2k)!}}{{4^k (k!)^2 }}}
} \\ {c_{2k + 1} = 0} \\ \end{array}}\right. ,
\]%
for $k=0,1,2,3...$, with 
\[
\tanh \left[ \theta (h>1) \right] =\frac{1-\gamma }{2h-1-\gamma }.
\]%
The analytical form of fidelity between $h$ and $h^{\prime }$ can be obtain
as
\[
F(h,h^{\prime })=\frac{\left(1-\tanh ^{2}\frac{\theta }{2}\right)^{1/4}\left(1-\tanh ^{2}%
\frac{\theta ^{\prime }}{2}\right)^{1/4}}{\sqrt{\left(1-\tanh \frac{\theta }{2}\tanh
\frac{\theta ^{\prime }}{2}\right)}},
\]%
where $\theta =\theta (h)$ and $\theta ^{\prime }=\theta (h^{\prime })$%
. Fidelity here is an intensive value, by taking Log, the quantum logarithmic fidelity is
\begin{eqnarray}
\ln F(h,h^{\prime }) &=&\frac{1}{4}\ln \left(1-\tanh ^{2}\frac{\theta }{2}\right)+\frac{%
1}{4}\mathrm{ln\left(1-\tanh ^{2}\frac{\theta ^{\prime }}{2}\right)}  \nonumber \\
&-&\frac{1}{2}\mathrm{ln]\left(1-\tanh \frac{\theta }{2}\tanh \frac{\theta
^{\prime }}{2}\right).}
\end{eqnarray}%
Therefore, agreeing with the numerical result, the logarithmic fidelity is
independent of the system size in the polarized phase. With the expression of fidelity, we recover the fidelity susceptibility in 
Ref. \cite{HMKwok08FS} that by taking the second derivative of fidelity between $h$ and $h+dh$
with respect to $dh$ for $dh$ approaching $0$. It gives the result of,
\[
-\left. \frac{d^{2}F(h,h+\delta h)}{d\delta h^{2}}\right\vert _{\delta
h\rightarrow 0}=\frac{(1-\gamma )^{2}}{32(h-1)^{2}(h-\gamma )^{2}}.
\]% 
\subsection{Symmetry-broken phase $(h <1)$}

In the following, we are going to derive the $N$ dependence of logarithmic
fidelity for $h<1$, In semi-classical treatment\cite{SDusuel05,HMKwok08FS}, the relation between the spin operator at $h=h_1$ and $h'=h_2$ is given by 

\begin{figure}[tbp]
\includegraphics[width=8.5cm]{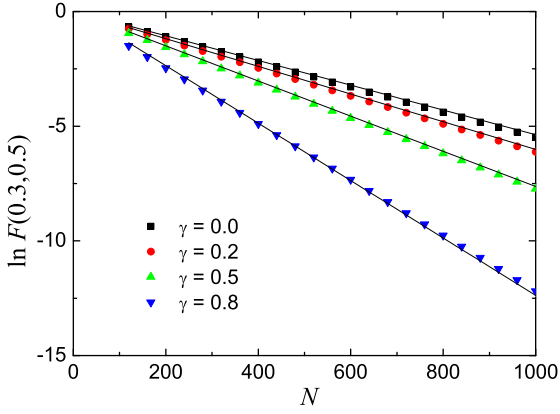}
\caption{(Color online)The dependence of logarithm of fidelity to the system size of the
LMG model in the symmetric-broken phase. Solid lines represent the
analytical solution and dots represent the numerical results.}
\label{fig:hsone}
\end{figure}

\[
\left\{ {\begin{array}{*{20}c} {\tilde S_{x,2} = \cos (\alpha _{12} )\tilde
S_{x,1} + \sin (\alpha _{12} )\tilde S_{z,1} } \\ {\tilde S_{z,2} = - \sin
(\alpha _{12} )\tilde S_{x,1} + \cos (\alpha _{12} )\tilde S_{z,1} } \\
\end{array}}\right. ,
\]%
where $\alpha _{i}=\arccos h_{i}$, $i=1$, $2$, $h_{i}$ representing two
different external field strength, $\alpha _{12}=\alpha _{1}-\alpha _{2}$
for a large system, $\tilde{S}$ is approximated in Holstein-Primakoff
representation that
\begin{eqnarray}
a_{2} &=&\frac{1}{2}\left[ {\cos (\alpha _{12})+1}\right] a_{1}+\frac{1}{2}%
\left[ {\cos (\alpha _{12})-1}\right] a_{1}^{\dag }  \nonumber \\
&+&\frac{1}{\sqrt{N}}\sin (\alpha _{12})\left( {\frac{N}{2}-a_{1}^{\dagger
}a_{1}}\right),   \nonumber \\
a_{2}^{\dagger } &=&\frac{1}{2}\left[ {\cos (\alpha _{12})-1}\right] a_{1}+%
\frac{1}{2}\left[ {\cos (\alpha _{12})+1}\right] a_{1}^{\dag }  \nonumber \\
&+&\frac{1}{\sqrt{N}}\sin (\alpha _{12})\left( {\frac{N}{2}-a_{1}^{\dag
}a_{1}}\right),
\end{eqnarray}%
where the $a^{\dag }$ and $a$ are bosonic creation and annihilation
operators. By same approach as the $h>1$ case, expanding the ground state in
terms of the eigenstate of $a_{1}^{\dagger }a_{1}$. Beware that both ground
states at $h_{1}$ and $h_{2}$ should be expanded in same set of basis (i.e. expand ground state at $h_2$ in terms of states at $h_1$). Thus we have
the expansion of ground state at different external field as
\begin{eqnarray}
\left\vert G\right\rangle _{1} &=&\left( {1-\tanh ^{2}\frac{{\theta _{1}}}{2}%
}\right) ^{\frac{1}{4}}\sum\limits_{m=0}^{\infty }{\tanh ^{m}\left( {\frac{{%
\theta _{1}}}{2}}\right) }\frac{{\sqrt{(2m)!}}}{{2^{m}m!}}\left\vert {2m}%
\right\rangle _{1}, \\
\left\vert G\right\rangle _{2} &=&\sum\limits_{k=0}^{\infty }{%
c_{k}\left\vert k\right\rangle _{1}},
\end{eqnarray}%
where
\begin{eqnarray}
c_{k} &=&-\frac{\Omega }{\Lambda }\cdot \frac{{\left( {\frac{N}{2}-k-1}%
\right) }}{{\sqrt{Nk}}}c_{k-1}-\frac{\Theta }{\Lambda }\sqrt{\frac{{k-1}}{k}}%
c_{k-2}, \\
c_{1} &=&-\frac{\Omega }{\Lambda }\cdot \frac{\sqrt{N}}{2}c_{o}, \\
\Omega  &=&\left( {\cosh \frac{{\theta _{2}}}{2}-\sinh \frac{{\theta _{2}}}{2%
}}\right) \sin (\alpha _{1}-\alpha _{2}), \\
\Lambda  &=&\frac{1}{2}\left\{ {\left[ {\cos (\alpha _{1}-\alpha _{2})+1}%
\right] \cosh \frac{{\theta _{2}}}{2}}\right\}   \nonumber \\
&-&\frac{1}{2}\left\{ {\left[ {\cos (\alpha _{1}-\alpha _{2})-1}\right]
\sinh \frac{{\theta _{2}}}{2}}\right\},  \\
\Theta  &=&\frac{1}{2}\left\{ {\left[ {\cos (\alpha _{1}-\alpha _{2})-1}%
\right] \cosh \frac{{\theta _{2}}}{2}}\right\}   \nonumber \\
&-&\frac{1}{2}\left\{ {\left[ {\cos (\alpha _{1}-\alpha _{2})+1}\right]
\sinh \frac{{\theta _{2}}}{2}}\right\} ,
\end{eqnarray}%
and,
\[
\tanh \left[ \theta _{i}(h<1)\right] =\frac{h_{i}^{2}-\gamma }{%
2-h_{i}^{2}-\gamma },
\]%
for $i=1$,$2$.

Now, we experienced in LMG model that, logarithm of fidelity is of different
dependence of system size in two phases. In other words, the .ogarithmic fidelity is
extensive in broken phase but intensive in symmetric phase. As we obtained in
the first page that the logarithm of fidelity in thermal states are always
proportional to the system size, we believe that the non-universal system
size dependence of logarithmic fidelity is a pure quantum effect.

\section{Scaling behaviour of logarithmic fidelity around the critical point}

\begin{figure}[tbp]
\includegraphics[width=8.5cm]{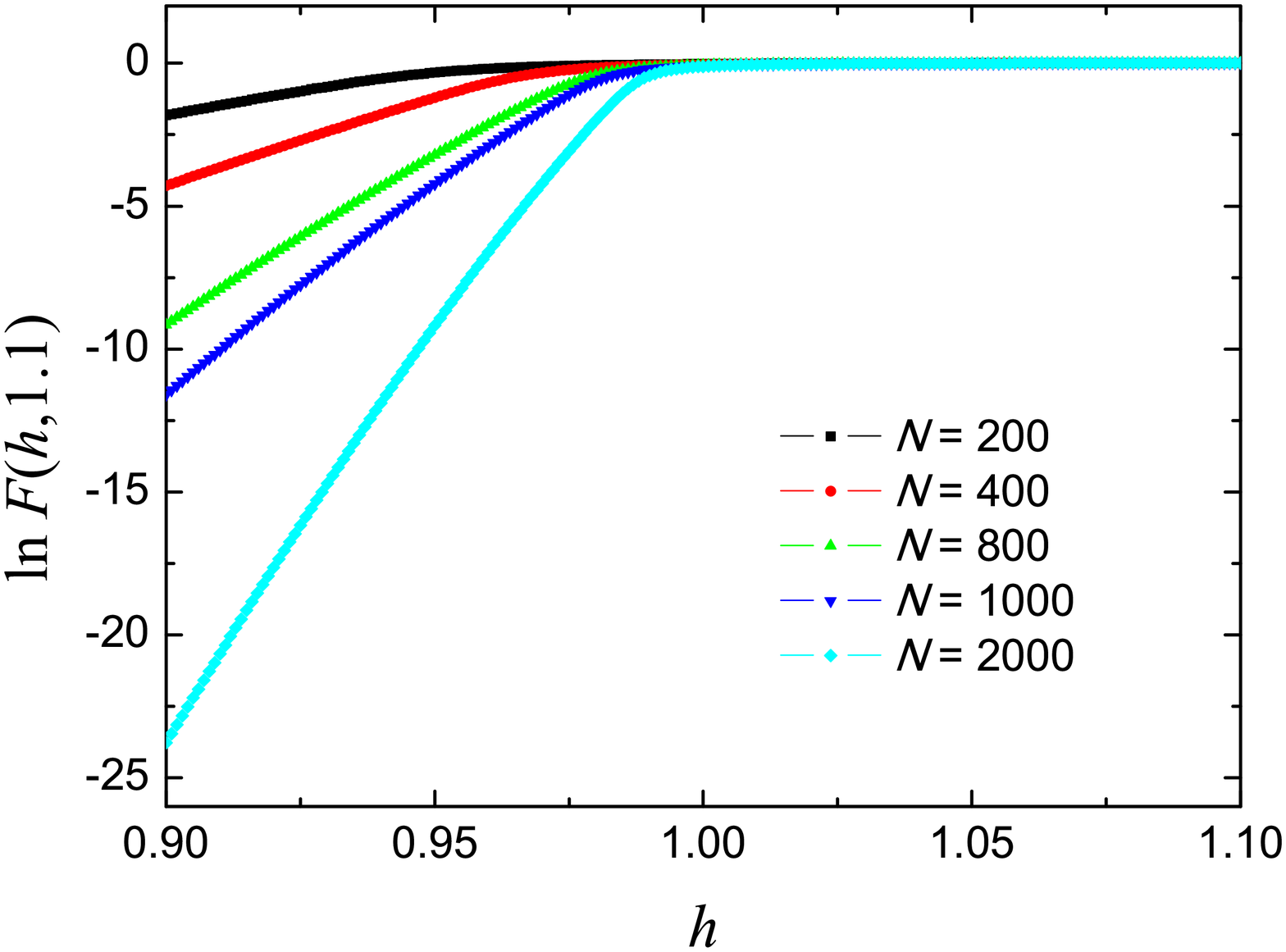}
\caption{(Color online)The logarithmic fidelity as a function of $h$ for various systems
sizes. Here the fixed point set at $h=$ 1.1.}
\label{fig:fig5}
\end{figure}

\begin{figure}[tbp]
\includegraphics[width=8.5cm]{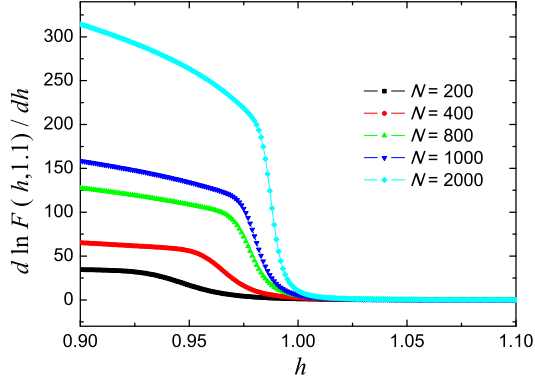}
\caption{(Color online)The first derivative of logarithmic fidelity for $h$ smaller than
one. Here the fixed point is set at $h=$ 1.1.}
\label{fig:fig6}
\end{figure}

\begin{figure}[tbp]
\includegraphics[width=8.5cm]{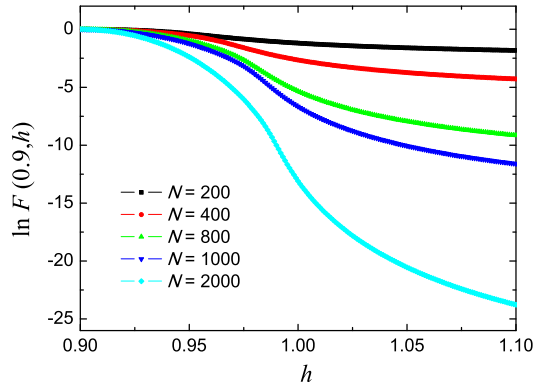}
\caption{(Color online)The logarithmic fidelity as a function of $h$ for various systems
sizes. Here the fixed point set at $h=$ 0.9.}
\label{fig:fig3}
\end{figure}

It is shown in Fig. \ref{fig:fig3} that for setting the fixed point in the
symmetry-broken phase, $\ln F$ experiences the extensive behavior across $h$%
. However, in Fig. \ref{fig:fig5} and Fig. \ref{fig:fig6}, which the fixed
point is set in the polarized phase $h=1.1$. For $h>1$, $\ln F$ is very
close to zero and exhibits itself intensively. However extensive behavior emerges once $h$ is smaller than the critical point $h=1$, despite finite size effects. Therefore, different scaling dependence is observed.
In the other point of view, in the symmetry-broken phase for infinite system
size, fidelity $F(h,h^{\prime })$ is zero unless $h=h^{\prime }$, or
\[
F(h,h^{\prime })=\delta( h,h^{\prime }).
\]%
This phenomenon is known as the Anderson orthogonal catastrophe. A tiny change
of $h$ makes a entirely orthogonal states. 

\begin{figure}[tbp]
\includegraphics[width=8.5cm]{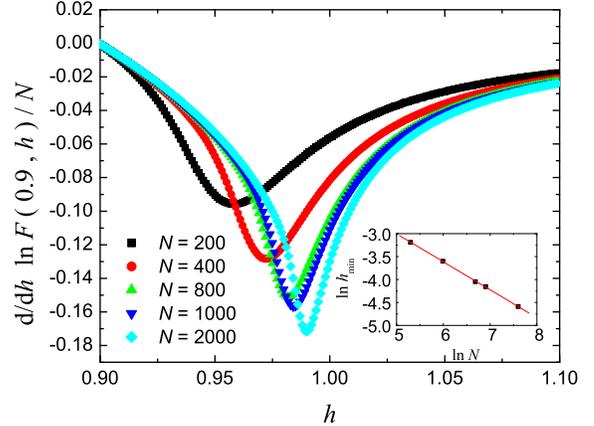}
\caption{(Color online)The first derivative of logarithmic fidelity, averaged due to system size. Fixed point was set at $h=$ 0.9, its minima
shows a nearly linear dependence with logarithmic system size(plotted in the
small graph), which shows that QPT occurs at $h_c=1$ for infinitly large system.}
\label{fig:fig4}
\end{figure}

\begin{figure}[tbp]
\includegraphics[width=8.5cm]{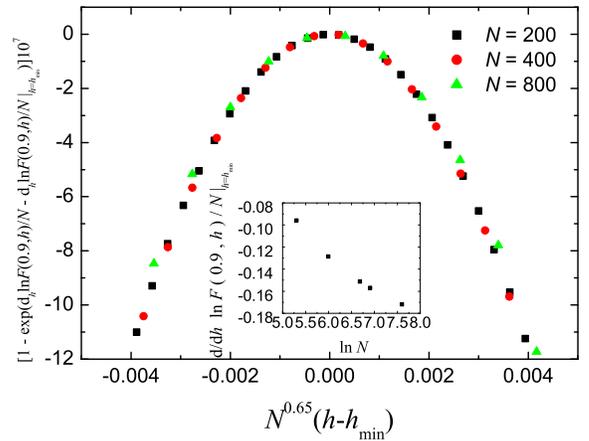}
\caption{(Color online)The finite size scaling analysis is performed. The critical
exponent is 0.65. The small graph shows the relation between the critical $h$
and system size.}
\label{fig:fig7}
\end{figure}

The first derivative of logarithmic fidelity in broken phase shows a minimum
occurs as quasi-critical point, $h_{\mathrm{min}}$, which approaches one as
\[
h_{\mathrm{m}}=1-1.0459N^{-0.6097},
\]
as shown in the small graph of Fig. \ref{fig:fig4}. It simply projects to $h_{c}=1$ as the critical point of LMG model. In Fig. \ref{fig:fig7} we examined the critical exponent of $\frac{d}{dh} \frac{\ln F (0.9,h)}{N}$. We analyzed the scaling function in the following form:
\[
1-\exp \left( \left. \frac{d}{dh}\frac{\ln F}{N}-\frac{d}{dh}\frac{\ln F}{N}%
\right\vert _{h=h_{\mathrm{min}}}\right) =f\left[ N^{\nu }(h-h_{\mathrm{min}%
})\right] ,
\]%
the parameter $\nu $ is the best adjusted when curves of different size overlap. It gives $\nu \sim
0.65$, which is close to the accepted value $\nu = \frac{2}{3}$.

The minima of first derivative of logarithmic fidelity shows slow
divergence. It is less than $N\ln N$ but greater than $\ln N$.

\section{Summary}

We first reviewed the classical logarithmic fidelity in phase transitions, and showed
the logarithmic fidelity is actually the Helmohlz Free Energy, which is an extensive quantity.
Then we analyzed the LMG model numerically, and discovered the quantum logarithmic fidelity
scales as $N$ in the broken phase and $N^0$ in the symmetric phase. Such a
change is not observed in classical systems, and we believe it is a quantum
effect.

This finding leaves a question to the definition of fidelity per site, where
``per site" may occasionally not hold. Recent papers concerning the
infinitesimal change of parameter in fidelity, taking its leading term, the
fidelity susceptibility shows similar behavior. In Ref. \cite{HMKwok08FS},
the authors gave analytic calculation of the fidelity susceptibility and
showed the different critical exponent in two phases of the LMG model. They
also confirmed with the scaling ansatz in which the critical exponent is
related to the scaling dependence of the system. On the other hand, for the
topological phase transition of the Kitaev model, the fidelity
susceptibility also exhibits different scaling dependence in two phases \cite%
{SYang08FS}. The difference is even more interesting, the fidelity
susceptibility scales like $L^2$ in one phase and $L^2 \ln{L}$ in another,
it changes from an extensive quantity to a superextensive quantity. In Ref.
\cite{SJGu08}, the extra $\ln{L}$ dependence is considered as the characteristic in the topological QPT.

\section{Acknowledgement}

This work was supported by the Earmarked Grant for Research from the
Research Grants Council of HKSAR, China (Project No. CUHK 400807).

\end{document}